\documentclass[aps,showpacs,twocolumn]{revtex4}
\usepackage{graphicx}
\usepackage{amsmath}
\usepackage{amsfonts}
\usepackage{amssymb}
\begin{document}
\newcommand{\be}{\begin{equation}}
\newcommand{\en}{\end{equation}}
\title{Bekenstein Bound and Spectral Geometry}
\author{Luis Alejandro Correa-Borbonet}\email{borbonet@cpd.ufmt.br}
\affiliation{Departamento de F\'\i sica \\ Universidade Federal de Mato Grosso,\\
Av. Fernando Corr\^{e}a da Costa, $s/n^o$-Bairro Coxip\'{o}\\
78060-900-Cuiab\'{a}-MT, Brazil}

\begin{abstract}
In this letter it is proposed to study the Bekenstein's $\xi(4)$  calculation of the $S/E$ bound for more general geometries. It is argued that,
using some relations among eigenvalues obtained in the context of Spectral Geometry,  it is possible to estimate $\xi(4)$  without an exact analytical
knowledge of the spectrum. Finally it is claimed that  isospectrality can define a class of domains with the same ratio $S/E$.
\end{abstract}
\pacs{02.70.Hm}

\maketitle
\section{Introduction.}
Since the pioneer works of Bekenstein\cite{bekenstein} and Hawking\cite{hawking}
about the gravitational entropy a substantial amount of work has been done trying to
understand the amazing connection between the area, a geometrical
quantity, and the entropy, a thermodynamic one. Within this context,
Bekenstein also proposed\cite{bekenstein1} the existence of a universal bound of
magnitude $2\pi R$ to the entropy-to-energy ratio $S/E$ of an
arbitrary system of effective radius $R$, or
\be
S/E \leq 2\pi
R.\label{eqn:bekbound}
\en
Originally, the bound was deduced by considering a gedanken
experiment of lowering the system into a black hole and demanding
this process to satisfy the genera\-lized second law of thermodynamics. On the
other hand, one expects that there must be a limit to the entropy that can be
placed in a system of finite size whose energy is limited. This is
suggested by the limited phase space available to the components
of such a system.

Besides that, Bekenstein himself proposed a explicit method to calculate the ratio $S/E$
for fields inside symmetric cavities in two dimensions like the square, the rectan\-gle and also in three dimension 
for fields inside the sphere,
the cube, etc\cite{bekenstein2}. This calculation was done using the known spectrum (eigenvalues) for these geo\-metries. Obviously the number
of examples was limi\-ted to the few cases where it is known the analytic form of the spectrum.

In this paper we will show some interesting results co\-ming from Spectral Geometry that allow to generalize the above mentioned Bekenstein method
to more general geo\-metries. Specifically, we will present some useful relations among the eigenvalues for generic domains that we will use to estimate
the $S/E$ ratio for a particular case.
Finally we will illustrate how the connections between the Bekenstein's proposal and Spectral Geometry help to set more clearly the reasons behind
the Bekenstein bound.

\section{Bekenstein approach}
In the work \cite{bekensschif} was shown that if the cavity
confining the system is circumscribed by a sphere of radius $R$
then the microcanonical  entropy $S(E)=ln \,\Omega(E)$ obeys
\be S/E \leq [24\xi(4)]^{1/4}\; ,\label{eqn:sbproof}
\en
where
$\xi(k)$ is the $\xi$-function
\be
\xi(k)=\sum_{i} g_{i}\,\omega^{-k}_{i}\label{eqn:spectrumd}
\en
for the sphere,
$\{\omega_{i}\}$ is the discrete one-particle energy with
zero-modes excluded and $g_{i}$ represents the degeneracy of the
$i$-th level. Since for the sphere we have $\xi(4)\sim R^{4}$ the
bound (\ref{eqn:bekbound}) follows from (\ref{eqn:sbproof})
provided $R^{-4}\xi(4)$ is appropriately bound from above. The
later was verified in \cite{bekensschif} for various types of free
fields satisfying Dirichlet or Neumann conditions.

For the sake of simplicity  we show the case of the scalar field
inside and sphere. The solutions of the scalar equation which are
harmonic in time may be found only for discrete eigenfrequencies
$\omega_{i}$ which arise from the eigenvalue problem defined by
\be
\bigtriangledown^{2}\phi=-\omega^{2}\phi \; ,\label{eqn:eigenv}
\en
together with the Dirichlet boundary conditions for $\phi$. In
this case the solutions are $j_{n,l}(\omega
r)Y_{lm}(\theta,\phi)$, where $j_{l}$ is the standard spherical
Bessel function of order $l$. The boundary conditions then demands
that $wR$ be a positive zero of $j_{l}$. Hence the spectrum is
\begin{eqnarray}
w_{nl}=j_{n,l}R^{-1}\;\; , \;\;n=1,2,...;\;\;\; l=0,1,....\;\; ,
\end{eqnarray}
where $j_{n,l}$ is the nth positive zero of $j_{l}(x)$, the
degene\-racy is $2l+1$. The lowest eigenfrequency is
$\omega_{10}=\pi/R$. With this at hand it is possible to calculate
the analytical approximation to $max(S/E)$(for $R=1$), that is
\be
\xi(4)^{1/4}_{sphere}=0.452 \;\; .
\en
Similar computations were done for different cavities in one, two
and three dimensions\cite{bekenstein2}.

Another output of the works\cite{bekenstein2},\cite{bekensschif}
was the proof of a local theorem on the $\xi$ function. A precise
statement of this is that as a given
cavity $S$ is deformed into another one $\sum$ entirely contained
within it, all the eigenvalues increase and, therefore, the
function $\xi$ is smaller for $\sum$.

Related to the previous result is the existence of a lower bound
for the lowest eigenvalue of the scalar field in an arbitrary
odd-shaped cavity $C$ that is circumscribed by a sphere of radius
$R$, i.e,
\be
\omega_{1}> \pi/R,
\en
where the right hand side of
the inequality is the first eigenvalue of the scalar field in the
sphere \cite{bekensschif}. Bekenstein obtained this result by applying the
Rayleigh-Ritz principle to the eigenvalue equation
(\ref{eqn:eigenv}).

\section{Scalar fields in general manifolds}
In general, for any geometry, the one particle spectrum is poorly known, so in these situations it is
not possible to calculate explicitly $\xi(4)$.  Fortunately, in the last years the mathematicians working in the area of Spectral Geo\-metry have shed
light on this problem obtaining inte\-resting results about the relation  among the eigenvalues \cite{ashbaugh},\cite{esposito}. Therefore in this section
we will review some of the main inequalities for
the eigenvalues of the Laplacian on bounded domains in Euclidean
space. Our attention will be focused in the Dirichlet
Laplacian or fixed membrane eigenvalue problem, i.e, the problem\cite{ashbaugh}
\begin{eqnarray}
-\triangle \,u & = & \lambda \, u \;\;\;\;
\mathtt{in}\;\;\;\Omega\subset\mathbb{R}^{n},\nonumber \\
             u & = & 0 \;\;\; \mathtt{on} \;\; \partial\Omega,
             \label{eqn:dirich1}
\end{eqnarray}
where $\Omega$ is a bounded domain in Euclidean space
$\mathbb{R}^{n}$ and $\partial \Omega$ is its boundary (To avoid confusions with the previous notation it is worthy to point out
that $\lambda_{i}=\omega^{2}_{i}$).\\
It is well known that this problem has a real and purely discrete
spectrum $\{\lambda_{i}\}^{\infty}_{i=1}$, satisfying,
\be 
0<\lambda_{1}<\lambda_{2}<\lambda_{3}<\dots \lambda_n \dots  \rightarrow
\infty \;\; .
\en
Here each eigenvalue is repeated according to its multiplicity.

In general, to solve the problem (\ref{eqn:dirich1}) is a difficult
task and exact analytical solutions can be obtained just for some
domains. However, there are some techniques that allows to
obtain information about the bounds and relations satisfied by the
eigenvalues. In the following lines we will present some of these
results. Initially we show the  Rayleigh-Ritz inequality, which gives a
simple way to bound eigenvalues from above based on trial
functions, i.e

\be
\lambda_{1}(\Omega)= \inf_{\varphi \in D(-\triangle)}
\frac{\int_{\Omega} \varphi(-\triangle \varphi)
}{\int_{\Omega}\varphi^{2}},
\en
where $\varphi$ is a real trial function in the domain of
$-\triangle$. This kind of bound can be extended to higher
eigenvalues by imposing orthogonality conditions on the class of
trial functions used.

One of the earliest isoperimetric inequalities for an eigenvalue
is certainly that for the first eigenvalue of the Dirichlet
Laplacian and takes the form:
\be \lambda_{1}(\Omega)\geq \lambda_{1}(\Omega^{*}) \;\;\;\;
\mathtt{for} \;\;\;\Omega\subset\mathbb{R}^{n}\;\;,\label{eqn:faberk}
\en
with equality if and only if $\Omega$ is a
ball, i.e.,$\Omega=\Omega^{*}$. This is known as the Faber-Krahn
Inequality.

The next isoperimetric result is for the quotient between the
first two eigenvalues $\lambda_{1}$ and $\lambda_{2}$. In $1955$
and $1956$ Payne, Polya and Weinbeger(henceforth PPW) proved that\cite{PPW2}
\be
\left.\frac{\lambda_{2}}{\lambda_{1}}\right|_{\Omega}\leq 3 \;\;\;\; \mathtt{for}
\;\;\;\Omega\subset\mathbb{R}^{2}\;\; ,
\en
and conjectured that
\be
\left.\frac{\lambda_{2}}{\lambda_{1}}\right|_{\Omega}\leq \left.
\frac{\lambda_{2}}{\lambda_{1}}\right|_{\mathtt{disk}}
=\frac{j^{2}_{1,1}}{j^{2}_{0,1}}\thickapprox 2.5387 \;\; ,
\en
with equality if and only if $\Omega$ is a disk, $j_{p,k}$ denotes
the $k^{th}$ positive zero of the Bessel function. The analogous
results for higher dimensions are
\be
\left.\frac{\lambda_{2}}{\lambda_{1}} \right|_{\Omega}\leq 1+\frac{4}{n} \;\;\;\;
\mathtt{for} \;\;\;\Omega\subset\mathbb{R}^{n} \;\; ,
\en
and the PPW conjecture
\be
\left.\frac{\lambda_{2}}{\lambda_{1}}\right|_{\Omega}\leq \left.
\frac{\lambda_{2}}{\lambda_{1}}\right|_{\mathtt{n-ball}}
=\frac{j^{2}_{n/2,1}}{j^{2}_{n/2-1,1}}\;\; ,\label{eqn:relation}
\en
with equality if and only if $\Omega$ is an $n$-ball. This PPW
conjecture was proved in the work \cite{PPw}.

The search for relations between the eigenvalues was extended to
higher eigenvalues in the form of universal inequalities and in
$1955$ Payne, Polya and Weinbeger proved also that
\be \lambda_{m+1}-\lambda_{m}\leq
\frac{2}{m}\sum^{m}_{i=1}\,\lambda_{i}, \;\;\; m=1,2,\dots . \en
 for $\Omega\subset\mathbb{R}^{2}$. This result extends to
$\Omega\subset\mathbb{R}^{n}$ as
\be
\lambda_{m+1}-\lambda_{m}\leq
\frac{4}{mn}\sum^{m}_{i=1}\,\lambda_{i},
\;\;m=1,2,\dots \;\; .\label{eqn:universal}
\en
The inequality
(\ref{eqn:universal}) is called a universal inequality because it
applies to all domains $\Omega\subset\mathbb{R}^{n}$,
"universally"\cite{ashbaugh2}.

A stronger inequality was derived by Hile and
Protter\cite{hiprotter} who proved that
\be
\sum^{m}_{i=1}\frac{\lambda_{i}}{\lambda_{m+1}-\lambda_{i}}\geq
\frac{mn}{4} \;\;\;\mathtt{for}
\;\;\;m=1,2,\dots \;\;.\label{eqn:hilprott}
\en
Note that (\ref{eqn:hilprott}) implies (\ref{eqn:universal}),
since we can replace the $\lambda_{i}$ in the denominator of
(\ref{eqn:hilprott}) by $\lambda_{m}$ to obtain
(\ref{eqn:universal}).

More recently, Yang \cite{yang} derived the inequality
\begin{eqnarray}
\sum^{m}_{i=1}(\lambda_{m+1}-\lambda_{i})\left(\lambda_{m+1}-\left(1+\frac{4}{n}\right)\lambda_{i}
\right)\leq 0  \label{eqn:yang1}\\ \quad \quad \quad \quad \quad \mathtt{for} \;\;\;m=1,2,\dots . \nonumber
\end{eqnarray}
This inequality will be referred as Yang's first inequality to
distinguish it from a simpler inequality implied by it(to be
called Yang's second inequality). Inequality (\ref{eqn:yang1}) is
an implicit bound for $\lambda_{m+1}$, but an explicit bound can
be derive from it by observing that its left hand side is just
quadratic in $\lambda_{m+1}$. Therefore, taking the larger root
and using the Cauchy-Schwarz inequality allow us to arrive at the
Yang's second inequality
\be
\lambda_{m+1}\leq
\left(1+\frac{4}{n}\right)\frac{1}{m}\sum^{m}_{i=1}\,\lambda_{i} .\label{eqn:yang2}
\en
This inequality is clearly stronger than the PPW inequality, since
it results from replacing the $\lambda_{m}$ by the average of the
first $m$ eigenvalues and $\lambda_{m}$ is certainly larger than
or equal than the average. Thus, we conclude that both of Yang's
inequalities are stronger than the PPW inequality. On the other
hand it can be proved also that the Yang inequality is stronger than
the HP inequality\cite{ashbaugh2}. This lead us to the following
relations
\be
\mathtt{Yang}\; 1 \Longrightarrow \mathtt{Yang}\; 2
\Longrightarrow \mathtt{Hile-Protter} \Longrightarrow \mathtt{PPW} \;\; .
\en
Although other interesting relations among the eigenvalues can
be found in the literature those showed before are useful
enough for our purposes.
\subsection{Estimation of $\xi(4)$ for scalar fields in a deformed spherical cavity}
In the first section was mentioned that the calculation of $\xi(4)$ for
the scalar fields  was done for various symme\-tric domains. Our purpose here is
to give an estimate of this quantity for a domain obtained from an
slight deformation of an sphere. In order to do that we will do
some plausible considerations and use the relations among the
eingevalues presented in the previous section.

Our first consideration is that in this domain there is not
degeneracy, therefore $g_{i}=1$. This can be seen as a consequence
of the deformation of the sphere that breaks all its symmetries.
Now, taking into account the Faber-Krahn
inequality(\ref{eqn:faberk}) we assume, for example, that the first eigenvalue
of the domain under study is $1$ percent bigger than the first
eigenvalue of the corresponding spherical problem, i.e,
$j_{1/2,1}=\pi$. In order to get the second eigenvalue we could
use the PPW inequality (\ref{eqn:relation})
\be
\frac{\lambda_{2}}{\lambda_{1}}\leq \left.
\frac{\lambda_{2}}{\lambda_{1}}\right|_{\mathtt{3-ball}}
=\frac{j^{2}_{3/2,1}}{j^{2}_{1/2,1}}=2.04484.
\en
To keep this inequality safe the second eigenvalue of the
spherical problem can not be modified in an amount equal or
superior to $1$ percent. Therefore we assume a modification of
$0.9$ percent. Then the quotient between the two first eigenvalues
is
\be
\frac{\lambda_{2}}{\lambda_{1}}=2.04080< 2.04484.
\en

For the higher eigenvalues we will use a modification of the
Yang's second inequality(\ref{eqn:yang2}). Actually we modify 
the relation among the eigenvalues substituting the
factor $(1+\frac{4}{3})$ by $2.04080$. Therefore
\begin{eqnarray}
\lambda_{m+1}=\left(2.04080\right)\frac{1}{m}\sum^{m}_{i=1}
\lambda_{i}.
\end{eqnarray}
Using these assumptions and relations we are ready to calculate
$\xi(4)$ for this domain, giving
\be
\xi(4)^{1/4}_{dom}=\left(\sum_{i} \frac{1}{\lambda^{2}_{i}}\right)^{1/4}=0.3536\;\; .
\en
This value is about $78$ percent of the value obtained for the sphere.
At this point would be interesting to note that if we calculate again $\xi(4)$ for 
the sphere and we neglect the degeneracies $2l+1$ the result is
\be
\xi(4)^{1/4}_{sphere}=0.3586 \;\;\;\mathtt{for} \;\;\;g_{i}=1\; .
\en
Therefore $\xi(4)^{1/4}_{dom}$ would
be $98.59$ percent of $\xi(4)^{1/4}_{sphere}$ (assuming $g_{i}=1$). Consequently,
we can conclude that in the case of a slight deformation of the
cavity, from the spherical symmetry, the main cause in the
decrease of $\xi(4)$ is due to the lost of degeneracies in the
eigenvalues.

Obviously that, for a real case, we would need to know a clear
relation between the degree of deformation of the geometry and the
change in the eigenvalues. In this case the crucial point would be
to do the best estimation of the first eigenvalue. Doing that we
could obtain an acceptable estimation of $\xi(4)$.


\section{On hearing the shape of a drum and isospectrality}
At this point it is not difficult to imagine that the knowledge of the
spectrum of a determined domain can help us to gain essential
information of the system. Already in $1911$ Herman Weyl proved
that the area of a plane domain is determined by its
spectrum\cite{weyl}. Some years later the Swedish mathematician
Ake Pleijel also proved that it is possible to obtain the length of the boundary
of the domain \cite{pleijel} from the spectrum as well. These relations between the
spectrum and the geome\-trical properties of the domain can be shown explicitly using the trace of the heat kernel, ie,
\be
 Z(t)=\sum^{\infty}_{n=1}e^{-\lambda_{n}t},
\en
where ${\{\lambda_{n}\}}$ are the eigenvalues of the Laplace operator. If the domain $M$ has a smooth boundary, $Z(t)$ has an asymptotic
expansion for a small positive $t$, given by the Minakshisundarum-Pleijel formula,
\be
Z(t)=\frac{1}{4\pi t} \sum^{\infty}_{k=0} D_{k}t^{k/2},
\en
where coefficients $D_{k}$ reflect the geometric nature of the domain $M$. Particularly,
\begin{eqnarray}
D_{0} & = & Area(M), \\
D_{1} & = & -\frac{\sqrt{\pi}}{2}  Length(\partial M).
\end{eqnarray}
These interesting results led to the
speculation that perhaps the shape of a plane domain(or more
generally, of a Riemannian manifold) is audible. It is worthy to
remember that if $M$ is a domain in the Euclidean plane then the
Dirichlet eigenvalues of $\triangle$ are essentially the
frequencies produced by a drumhead shaped like $M$. In this line,
in a landmark paper\cite{kac}, Mark Kac posed the question "Can one
hear the shape of a drum?". In the case of a Riemannian manifold, the Kac's question can paraphrased as:
"Can one deduce the metric of the surface from the spectrum?". Until the moment the answer to this question is not known in sufficient detail. An affirmative answer is known to hold for several classes of surfaces and domains\cite{gutkin},\cite{zelditch}. However, this is not always true. One of the first examples to the negative
is due to Milnor who proposed in $1964$ two flat tori in
$\mathbb{R}^{16}$, which he proved to be isospectral but not
isometric. Since then, many other pairs of isospectral(counting multiplicities) yet not
isometric systems were found. A general method for constructing
isospectral, non-isometric manifolds has been designed by
Sunada\cite{sunada}. Despite these advances in se\-veral dimensions the problem for plane regions remained open until 1991, when Carolyn Gordon, David Webb, and Scott Wolpert found examples of distinct plane "drums" which "sound" the same. Lately this was confirmed experimentally by the work of Sridhar and Kudrolli \cite{sridhar}. In the experiments they employed thin microwave cavities shaped in the form of two different domains known to be isospectral. Specifically, they verified the equality of at least $54$ of the measured low-lying eigenvalues to a few parts in $10^{4}$. On the other hand Driscoll \cite{driscoll} showed a method to evaluate numerically the eigenvalues of poly\-gonal regions.

On the light of the results presented above we could concluded
that isospectral domains, in the case of scalar fields, have the
same relation $S/E$. In other words, isospectrality allows to define a 
class of systems with the same ratio $S/E$. That is obviously clear from the form of
$\xi(4)$. On the other hand geometric constraints are forced on isospectral manifolds and 
this fact could su\-ggest that these domains have the same effective radius. Therefore we
can concluded that these results coming from the field of Spectral
Geometry support strongly the Bekenstein proposal. This conclusion could be reinforced by some 
results, recently found, that relate  Information Theory and Spectral Geometry\cite{kempf}.

\begin{acknowledgments}
I would like to thank I. Cabrera Carnero for useful discussions and suggestions. This work has been
 supported by CNPq-FAPEMAT/UFMT.
\end{acknowledgments}

\end{document}